\documentclass{article}
\begin{document}

\title{Teleportation via thermally entangled state of a two-qubit Heisenberg XX chain}
\author{Ye Yeo}
\maketitle

\begin{center}
{\it Centre for Mathematical Sciences, Wilberforce Road, Cambridge CB3 0WB, United Kingdom}
\end{center}

\begin{abstract}
We find that quantum teleportation, using the thermally entangled state of two-qubit Heisenberg XX chain as a resource, with fidelity better than any classical communication protocol is possible.  However, a thermal state with a greater amount of thermal entanglement does not necessarily yield better fidelity.  It depends on the amount of mixing between the separable state and maximally entangled state in the spectra of the two-qubit Heisenberg XX model.
\end{abstract}

The linearity of quantum mechanics allows building of superposition states of composite system
$S_{AB}$
that cannot be written as products of states of each subsystem
$(S_A$ and $S_B)$.
Such states are called entangled.  States which are not entangled are referred to as separable states.  An entangled composite system gives rise to nonlocal correlation between its subsystems that does not exist classically.  This nonlocal property enables the uses of local quantum operations and classical communication to transmit an unknown state
$|\psi\rangle$
via a shared pair of entangled particles, with fidelity better than any classical communication protocol \cite{Bennett,Popescu,Horodecki}.  The standard teleportation protocol
$P_0$
uses Bell measurements and Pauli rotations.  Standard teleportation with an arbitrary mixed state resource
$\rho_{AB}$
is equivalent to a generalized depolarizng channel
$\Lambda_{P_0}(\rho_{AB})$
with probabilites given by the maximally entangled components of the resource \cite{Bose}.  The fidelity for
$P_0$
and
$\rho_{AB}$
is defined by averaging
$\langle \psi| \left[ \Lambda_{P_0} \left( \rho_{AB} \right) |\psi \rangle \langle \psi| \right] |\psi \rangle$
over all possible
$|\psi\rangle$.
Quantum teleportation can thus serve as an operational test of the presence and strength of entanglement in
$\rho_{AB}$.\\

Recently, the presence of entanglement in condensed-matter systems at finite temperatures has been investigated by a number of authors (see, e.g., \cite{Nielsen} and references therein).  The state of a typical condensed-matter system at thermal equilibrium (temperature $T$) is
$\rho = e^{-\beta H}/Z$
where $H$ is the Hamiltonian,
$Z = tr e^{-\beta H}$
is the partition function, and
$\beta = 1/kT$
where $k$ is Boltzmann's constant.  The entanglement associated with the thermal state $\rho$ is referred to as the thermal entanglement \cite{Arnesen}.  The bulk of these investigations concentrated on the quantification of thermal entanglement and how this quantity changes with temperature $T$ and external magnetic field $B_m$.  This leaves open the question if thermal entanglement is useful for quantum information tasks like quantum teleportation, which is not only relevant to quantum communication between two distant parties but also to quantum computation.  Quantum teleportation is a universal computational primitive \cite{Chuang}.  With proposals like the one-way quantum computer \cite{Briegel, Browne}, it is important to find out the answer to this question.\\

In this paper, motivated by \cite{Briegel, Browne}, we take the first step by considering quantum teleportation in the two-qubit Heisenberg XX model \cite{Wang}.  We find that although quantum teleportation with fidelity better than any classical communication protocol is possible, the amount of nonzero thermal entanglement does not guarantee this.  In fact, we could have a more entangled thermal state not achieving a better fidelity than a less entangled one.\\

The Hamiltonain $H$ for a two-qubit Heisenberg XX chain in an external magnetic field $B_m$ along the $z$ axis is
\begin{equation}
H = \frac{1}{2} J   \left(\sigma^1_A \otimes \sigma^1_B +
                          \sigma^2_A \otimes \sigma^2_B \right)+
    \frac{1}{2} B_m \left(\sigma^3_A \otimes \sigma^0_B +
                          \sigma^0_A \otimes \sigma^3_B \right)
\end{equation}
where
$\sigma^0_{\alpha}$
is the identity matrix and
$\sigma^i_{\alpha} (i=1,2,3)$
are the Pauli matrices at site
$\alpha = A, B$.
$J$ is real coupling constant for the spin interaction.  The chain is said to be antiferromagnetic for $J>0$ and ferromagnetic for $J<0$.  The eigenvalues and eigenvectors of $H$ are given by
$H |00\rangle = B_m |00\rangle$,
$H |\Psi^{\pm}\rangle = \pm J |\Psi^{\pm}\rangle$ and
$H |11\rangle = -B_m |11\rangle$, where
$|\Psi^{\pm}\rangle = \frac{1}{\sqrt{2}} (|01\rangle \pm |10\rangle)$.
For the system in equilibrium at temperature $T$, the density operator is
\begin{equation}
\rho_{AB} = \frac{1}{Z}
\left[
e^{-\beta B_m} |00\rangle \langle 00| +
e^{-\beta J} |\Psi^+\rangle \langle\Psi^+| +
e^{\beta J} |\Psi^-\rangle \langle\Psi^-| +
e^{\beta B_m} |11\rangle \langle 11|
\right]
\end{equation}
where the partition function
$Z = 2\cosh\beta B_m + 2 \cosh\beta J$,
the Boltzmann's constant $k \equiv 1$ from hereon and
$\beta = 1/T$.
To quantify the amount of entanglement associated with $\rho_{AB}$, we consider the concurrence \cite{Wootters, Hill},
$C = \max\{ \lambda_1 - \lambda_2 - \lambda_3 - \lambda_4, 0\}$ where
$\lambda_k (k = 1,2,3,4)$
are the square roots of the eigenvalues in decreasing order of magnitude of the spin-flipped density matrix operator
$R = \rho_{AB} (\sigma^2 \otimes \sigma^2) \rho^{\ast}_{AB} (\sigma^2 \otimes \sigma^2)$,
where the asterisk indicates complex conjugation.  After some straightforward algebra, the thermal concurrence is
\begin{equation}
C(\rho_{AB}) = \max\left\{
\frac{|\sinh\beta J| - 1}{\cosh\beta B_m + \cosh\beta J}, 0
\right\}
\end{equation}
The concurrence is invariant under the substitutions
$B_m \rightarrow -B_m$ and
$J \rightarrow -J$.
The latter indicates that the entanglement is the same for the antiferromagnetic and ferromagnetic cases.  We thus restrict our considerations to
$B_m > 0$ and $J>0$.
Notice that the critical temperature
$T_{critical} \approx 1.13459 J$,
beyond which the thermal concurrence is zero, is independent of the magnetic field $B_m$.\\

Now we look at the standard teleportation protocol $P_0$, using the above two qubit mixed state $\rho_{AB}$ as a resource.  We consider as input a qubit in an arbitrary pure state
$|\psi\rangle = \cos\frac{\theta}{2} |00\rangle +
                e^{i\phi}\sin\frac{\theta}{2} |11\rangle$
$(0 \le \theta \le \pi, 0 \le \phi \le 2\pi)$.
The output state is then given by \cite{Bose},
\begin{equation}
\Lambda_{P_0}(\rho_{AB}) |\psi\rangle \langle\psi| =
\sum_{j = 0}^3 tr(E^j \rho_{AB})
\sigma^j_A |\psi\rangle \langle\psi| \sigma^j_A
\end{equation}
where
$E^0 = |\Psi^-\rangle \langle\Psi^-|$,
$E^1 = |\Phi^-\rangle \langle\Phi^-|$,
$E^2 = |\Phi^+\rangle \langle\Phi^+|$,
$E^3 = |\Psi^+\rangle \langle\Psi^+|$, and
$|\Phi^{\pm}\rangle = \frac{1}{\sqrt{2}} (|00\rangle + |11\rangle)$.
It follows that
\begin{equation}
\langle \psi| \left[ \Lambda_{P_0} \left( \rho_{AB} \right) |\psi \rangle \langle \psi| \right] |\psi \rangle =
\frac{2\sin^2\theta\cosh\beta B_m + (3 + \cos 2\theta)\cosh\beta J + 2\sin^2\theta\sinh\beta J}
{4(\cosh\beta B_m + \cosh\beta J)}
\end{equation}
and averaging over all possible input states $|\psi\rangle$ we obtain the fidelity of the teleportation
\begin{equation}
\frac{\cosh\beta B_m + 2\cosh\beta J + \sinh\beta J}{3(\cosh\beta B_m + \cosh\beta J)} =
\frac{2\frac{e^{\beta J}}{Z} + 1}{3}
\end{equation}
This is the maximal fidelity achievable from $\rho_{AB}$ in the standard teleportation scheme $P_0$ \cite{Horodecki}.  In order to transmit $|\psi\rangle$ with fidelity better than any classical communication protocol, we require (6) to be strictly greater than $\frac{2}{3}$.  In other words, we require
\begin{equation}
\sinh\beta J > \cosh\beta B_m
\end{equation}
and hence $B_m < J$.  The `critical' temperature $T^{(m)}_{critical}$ beyond which the performance is worst than what classical communication protocol can offer, is clearly dependent on the magnetic field $B_m$.  As shown in Table I, $T^{(m)}_{critical}$ decreases with increase in $B_m$ and they are all less than $T_{critical}\approx 1.13459J$.  This means that with increasing $B_m$ we have an increasing range of nonzero thermal entanglement which is however not `suitable' for quantum teleportation.  In fact, when $B_m \ge J$, we have nonzero thermal entanglement as long as the temperature $T$ of the system is less than $T_{critical}$.  Therefore, the thermal concurrence (3) only describes certain aspects of the thermal entanglement.  In particular, it does not indicate the `quality' of thermal entanglement for quantum teleportation.  Physically, one could understand the phenomenon by looking at the thermal density operator (2).  When $B_m \ge J$, one could attribute the cause of the poor quality of thermal entanglement to the fact that there is now a comparable or greater proportion of separable $|11\rangle\langle 11|$ than the maximally entangled $|\Psi^-\rangle\langle\Psi^-|$.  This reasoning could be carried over to the case when $B_m < J$ since (7) implies
$e^{\beta J}/Z > e^{\beta B_m}/Z$, and as long as $T<T^{(m)}_{critical}$ we have (7).\\

In conclusion, the condition (7) for the thermal quantum channel (2) to teleport states with fidelity strictly greater than $\frac{2}{3}$ is expectedly stronger than that for thermal concurrence (3) to be nonzero.  However, it does not rule out the possibility of using thermally entangled states to teleport quantum states with high fidelity.  It would be interesting to see how the entanglement of distillation \cite{Distil} depends on the temperature $T$ and magnetic field $B_m$.  The entanglement of distillation is the number of maximally entangled pairs that can be purified from a given state.  One would intuitively expect it to vary in the same fashion as the fidelity.  Unfortunately, we do not have a closed analytic expression for the entanglement of distillation.

\paragraph{Acknowledgements}
I would like to thank Yuri Suhov, Andrew Skeen and Suguru Furuta for useful discussions.  This publication is an output from project activity funded by The Cambridge MIT Institute Limited (``CMI'').  CMI is funded in part by the United Kingdom Government.  The activity was carried out for CMI by the University of Cambridge and Massachusetts Institute of Technology.  CMI can accept no responsibility for any information provided or views expressed.

\newpage
\begin{table}
\begin{tabular}{|c|c|l|}
\hline
$\eta$ & $T^{(m)}_{critical}$ & $C_r$ \\ \hline
0.1 & 1.13105$J$ & 0.00161554 \\ \hline
0.2 & 1.12029$J$ & 0.00654425 \\ \hline
0.3 & 1.10193$J$ & 0.0150472 \\ \hline
0.4 & 1.07525$J$ & 0.0276166 \\ \hline
0.5 & 1.03904$J$ & 0.045085 \\ \hline
0.6 & 0.991262$J$ & 0.068864 \\ \hline
0.7 & 0.928278$J$ & 0.101495 \\ \hline
0.8 & 0.842666$J$ & 0.148196 \\ \hline
0.9 & 0.714112$J$ & 0.223103 \\ \hline
\end{tabular} \\
Table I: \quad
The `critical' temperature $T^{(m)}_{critical}$ is function of both $J$ and $B_m = \eta J, 0 < \eta < 1$.  The `residual' concurrence $C_r$ gives the amount of thermal entanglement below which does not yield teleportation fidelity better than any classical communication protocol.
\par
\end{table}

\end{document}